\documentclass[%
superscriptaddress,
nofootinbib,
nobibnotes,
amsmath,amssymb,
pra,
]{revtex4-2}

\usepackage{graphicx}

\usepackage[utf8]{inputenc}
\usepackage{multirow}
\usepackage{xcolor}
\usepackage{hyperref}
\hypersetup{colorlinks=true,linkcolor=blue,anchorcolor=blue,citecolor=magenta,filecolor=blue,urlcolor=blue,bookmarksnumbered=true}
\usepackage[flushleft]{threeparttable}
\usepackage{mathtools}

\usepackage{stackrel}

\usepackage{dsfont}
\usepackage{amsthm}

\usepackage{enumitem}
\setlist[itemize]{leftmargin=10mm}

\usepackage{braket}

\begin{document}


\title{Systematic input scheme of many-boson Hamiltonians with applications to the two-dimensional $\phi ^4$ theory} 

\author{Weijie Du}
\email[Email:]{\ duweigy@gmail.com}
\affiliation{Department of Physics and Astronomy, Iowa State University, Ames, Iowa 50010, USA}

\author{James P. Vary}
\affiliation{Department of Physics and Astronomy, Iowa State University, Ames, Iowa 50010, USA}

\date{\today}

\begin{abstract}

We develop a novel, systematic input scheme for many-boson Hamiltonians in order to solve  field theory problems within the light-front Hamiltonian formalism via quantum computing. 
We present our discussion of this input scheme based on the light-front Hamiltonian of the two-dimensional $\phi ^4$ theory.
In our input scheme, we employ a set of quantum registers, where each register encodes the occupation of a distinct boson mode as binaries.
We squeeze the boson operators of each mode and present the Hamiltonian in terms of unique combinations of the squeezed boson operators.
We design the circuit modules for these unique combinations. 
Based on these circuit modules, we block encode the many-boson Hamiltonian utilizing the idea of quantum walk.
For demonstration purposes, we present the spectral calculations of the Hamiltonian utilizing the hybrid quantum-classical symmetry-adapted quantum Krylov subspace diagonalization algorithm based on our input scheme, where the quantum computations are performed with the IBM Qiskit quantum simulator. 
The results of the hybrid calculations agree with exact results.

\end{abstract}

\maketitle

\section{Introduction}

Quantum field theory (QFT) reconciles quantum mechanics and special relativity and it has important applications in many fundamental areas of physics \cite{peskin2018introduction}. Nonperturbative, {\it ab initio} structure and dynamics solutions of QFT are challenging on classical computers due to the nature of quantum many-body problems \cite{Preskill2018quantumcomputingin}. Quantum computing hardwares and algorithms hold the promise to address such numerical challenges \cite{feynman1982simulating}. 

Although, it is not currently possible to solve realistic problems on quantum computers in a way that outcompetes the best classical computers, it is an important and challenging task to develop new quantum algorithms that are suitable for solving QFT problems via {\it ab initio} methods on future fault-tolerant quantum computers. Following pioneering work by Jordan, Lee and Preskill \cite{Jordan_2012}, many quantum algorithms have been proposed for treating the QFT problems via the lattice-based Hamiltonian formalism \cite{PhysRevA.98.032331,PhysRevA.99.052335,PhysRevD.101.074512,Rhodes:2024zbr,bauer2022quantum,Bauer:2023qgm,Muller:2023nnk,Bauer:2021gek,Ciavarella:2024fzw,DAndrea:2023qnr,Hariprakash:2023tla,Davoudi:2024wyv,Watson:2023oov,Davoudi:2022xmb,Mueller:2022xbg,Rhodes:2024zbr,Rhodes2024unpub}.
In this cutting-edge area, major research focuses on reducing the error of digitization and latticing the continuous field onto qubit degrees of freedom as well as reducing the scaling of the computational resources \cite{PhysRevA.99.052335,Rhodes:2024zbr}. 

Fewer works employ the alternative promising light-front (LF) Hamiltonian formalism \cite{RevModPhys.21.392,Dirac_1950} in formulating the QFT problems in quantum computing. The LF Hamiltonian formalism has the attractive feature that the LF vacuum has a simple structure as the Fock vacuum is an exact eigenstate of the full normal-ordered LF Hamiltonian; this enables the systematic Fock-space expansion of the physical states \cite{BRODSKY1998299}. To date, the LF Hamiltonian formalism has witnessed successful initial applications in describing relativistic bound-state problems and dynamics in quantum electrodynamics, quantum chromodynamics, and chiral field theories on classical computers \cite{Vary:2009gt,Li:2018uif,Li:2020uhl,Li:2021zaw,Zhao:2013cma,Zhao:2013jia,Xu:2021wwj,Mondal:2019jdg,Lan:2019rba,Wiecki:2014ola,Zhao:2013cma,Zhao:2014xaa,Li:2017mlw,Li:2015zda,Qian:2020utg,Du:2019ips,Jia:2018ary,Honkanen:2010rc}. Many quantum algorithms have also been proposed to solve the structure and dynamics of hadronic systems \cite{Kreshchuk:2020aiq,Kreshchuk:2020dla,Kreshchuk:2020kcz,Qian:2021jxp,Barata:2023clv,Barata:2022wim,Yao:2022eqm,Wu:2024rod}.

However, a systematic Hamiltonian input scheme is a missing corner stone for a general-purpose quantum algorithm in computing the bound-state and scattering properties of the QFT problems based on the LF Hamiltonian formalism. An efficient input scheme is critical for large-scale calculations on future quantum computers, where the construction of the many-body Hamiltonian matrix is highly nontrivial. 
In Refs. \cite{Du:2023bpw,Du:2024zvr}, we proposed a systematic input scheme for second-quantized many-fermion Hamiltonians. 
This input scheme respects the symmetries of the many-fermion Hamiltonians and is flexible for incorporating the particle-number variations. 
Yet, it is still an important mission to develop the counterpart Hamiltonian input scheme for many-boson systems. 
One noticeable difficulty that hinders such a development comes from the Bose-Einstein statistics, where multiple bosons can occupy the same mode and such many-boson states require proper normalization \cite{merzbacher1998quantum}. This is in contrast to the many-fermion case where the Pauli exclusion principle holds and the many-fermion states do not encounter multiple-occupancy normalization factors. 

In this work, we present a novel and systematic input scheme for the many-boson Hamiltonians in quantum computing. We work in the second-quantized representation and introduce a new circuit representation of boson operators based on our encoding scheme of many-boson states. Compared to the prototype sparse-matrix-based Hamiltonian input models \cite{childs2010relationship,berry2012black} that work with the row and column indices of the Hamiltonian matrices in the framework of first quantization, our input model operates directly on the Fock states and offers a convenient approach to access the many-boson Hamiltonian matrix elements based on the Fock bases. 
Our input scheme does not require any specific oracles, which can be nontrivial in practical circuit constructions.
We remark that our input scheme respects the symmetries of the many-boson Hamiltonians by design. This enables a straightforward approach to prepare the many-boson states with desired symmetries on quantum computers.  
In this work, we provide an explicit and systematic circuit design for our input model. 

We present the discussion of our input scheme utilizing the two-dimensional $\phi ^4$ [$(\phi ^4)_2$] theory as it is one of the simplest interacting QFT problems, and illustrates the essential ideas and techniques. 
We address the difficulty in normalizing the many-boson state by introducing the squeezed (or scaled) boson operators. 
We also present explicit circuit designs of the squeezed boson operators as well as their combinations, which properly treats the normalization of many-boson states and encodes the normalization as the phases of qubit states. 
Based on these techniques, we show the construction of the quantum walk states \cite{PhysRevLett.102.180501,childs2010relationship,berry2012black} to block encode the many-boson Hamiltonian. 
Utilizing this input scheme, we can quantum compute the structure and dynamics of the many-boson system utilizing precise and efficient quantum algorithms \cite{PhysRevLett.118.010501,gilyen2019quantum,PRXQuantum.2.040203,PRXQuantum.3.040305}. 
For demonstration purposes, we present the spectral calculations of the LF Hamiltonian of the $(\phi ^4)_2$ theory implementing the hybrid quantum-classical symmetry-adapted (SA) quantum Krylov subspace diagonalization (QKSD) algorithm \cite{Du:2024zvr,Kirby_2023} based on our input scheme, where such spectral solutions shed lights on the phase transition of the $(\phi ^4)_2$ theory \cite{Harindranath:1987db,PhysRevD.105.016020}.

The arrangement of this paper is as follows. 
We present the $(\phi ^4)_2$ theory within the LF Hamiltonian formalism and discuss the basis choice in Sec. \ref{sec:theory}. 
In Sec. \ref{sec:Ham_Input}, we introduce our input scheme of the LF Hamiltonian of the $(\phi ^4)_2$ theory.
In Sec. \ref{sec:App_SA_QKSD}, we present the hybrid quantum-classical SA-QKSD algorithm for spectral calculations based on our many-boson Hamiltonian input scheme.
In Sec. \ref{sec:example}, we show a numerical example of spectral calculations utilizing the SA-QKSD algorithm.
We summarize in Sec. \ref{sec:summary_and_outlook}, where we also present an outlook.

\section{Theory}
\label{sec:theory}

\subsection{LF Hamiltonian of the $ (\phi ^4)_2$ theory}

In brief, we present necessary details of the LF Hamiltonian formalism of the $(\phi ^4)_2$ theory. 
Interested readers are referred to Refs. \cite{Harindranath:1987db,PhysRevD.105.016020} for more details. 

The Lagrangian density of the $ (\phi ^4 )_2$ theory is
\begin{equation}
	\mathcal{L} = \frac{1}{2} (\partial _{\mu} \phi \partial ^{\mu} \phi -m^2 \phi ^2) - \frac{\lambda }{4!} \phi ^4 ,
\end{equation}
where $m$ denotes the mass parameter while $\lambda $ is the coupling constant. One chooses $\lambda > 0$ and $m > 0$ so that the corresponding LF Hamiltonian is bounded and the vacuum state is the normal vacuum for small coupling in the symmetric phase of the theory. 
Following the discrete light-cone quantization (DLCQ) approach \cite{Pauli:1985ps, Pauli:1985pv, Eller:1986nt}, one obtains the total longitudinal momentum $ K$ and the LF Hamiltonian $ H $. These are given by Eqs. (8-11) in Ref. \cite{PhysRevD.105.016020} where the zero mode is neglected as
\begin{align}
	K =& \sum _{k=1} k a^{\dagger}_k a_k , \\
	H =& H_{1\rightarrow 1} + H_{2\rightarrow 2} + H_{3\rightarrow 1} + H_{1\rightarrow 3} ,
	\label{eq:original_Phi4_Hamiltonian}
\end{align}
with
\begin{align}
	H_{1\rightarrow 1} = & \sum _{k} \frac{1}{k} \Big[ m^2 + \underbrace{ \frac{\lambda}{4\pi} \frac{1}{2} \sum _l \frac{1}{l}   } \Big] a_k^{\dagger} a_k , \label{eq:H_1to1}  \\
	H_{2\rightarrow 2} = & \frac{\lambda}{4\pi} \sum _{k\leq l} \sum _{m \leq n} \frac{1}{N^2_{kl}} \frac{1}{N^2_{mn}} \frac{1}{\sqrt{klmn}} \Big[ a_k^{\dagger} a_l^{\dagger} a_m a_n \Big] \delta_{(m+n), (k+l)} , \label{eq:H22} \\
	H_{3\rightarrow 1} = & \frac{\lambda}{4\pi} \sum _k \sum _{l\leq m \leq n} \frac{1}{N^2_{lmn}} \frac{1}{\sqrt{klmn}} \Big[ a^{\dagger} _k a_l a_m a_n  \Big] \delta _{k, (m+n+l)} , \label{eq:H31} \\
	H_{1\rightarrow 3} = & \frac{\lambda}{4\pi} \sum _{k} \sum _{l\leq m\leq n} \frac{1}{N^2_{lmn}} \frac{1}{\sqrt{klmn}} \Big[ a^{\dagger} _n a^{\dagger} _m a^{\dagger} _l a_k \Big] \delta _{k, (m+n+l)} , \label{eq:H13}
\end{align}
with $ N_{kl} =1 $ for $k\neq l$ and $N_{kl} = \sqrt{2!}$ for $k=l$, and
\begin{align}
	N_{lmn } = 
	\begin{cases}
		1 , & \text{for} \ l \neq m \neq n , \\
		\sqrt{3!}, & \text{for} \ l = m = n , \\
		\sqrt{2!}, & \text{for any two of $l,m$, and $n$ are equal} .
	\end{cases}
\end{align}
The summations of $k$, $l$, $m$, and $n$ run over positive integers for symmetric boundary conditions which is the choice adopted here. 
Here, $H_{1\rightarrow 1}$ and $H_{2\rightarrow 2}$ are Hermitian and $H^{\dagger}_{3\rightarrow 1} = H_{1\rightarrow 3} $.
One notes that $H$ preserve the evenness and oddness of the particle number of the many-boson system.
The total longitudinal momentum $K$ is a conserved quantity in the LF Hamiltonian formalism as indicated by the Kronecker deltas in the above expressions and $[H, K]=0$.
According to $H$, one can compute the eigenvalues of the $(\phi ^4)_2$ theory. 
These eigenvalues are related to the mass spectrum of the invariant-mass operator 
\begin{equation}
	\hat{M}^2 = KH.
	\label{eq:m2_op}
\end{equation}
One can typically scale $H$ with $m^2$ and rewrite the interaction theory as a function of a single dimensionless variable $\lambda /m^2$ for convenience \cite{PhysRevD.105.016020}. With this scaling, $H$ has the dimension of [MeV]$^2$ and $K$ is dimensionless.

We define $a^{\dagger}_k$ and $a_k$ as the creation and annihilation operators for the mode with the longitudinal momentum $k$, respectively. 
These boson operators satisfy the commutation relations \cite{merzbacher1998quantum}
\begin{equation}
	[a_m,a_n] = [a^{\dagger}_m, a^{\dagger}_n] = 0 , \ [a_m , a^{\dagger}_n ] = \delta _{m,n} .
\end{equation}
For the LF Hamiltonian $H$ [Eq. \eqref{eq:original_Phi4_Hamiltonian}] retained in this work, we note that there are at most four boson operators in all the monomials.

We denote $\ket{k^{r_k}}$ to be the normalized many-boson state with $r_k$ bosons being in the mode $k$. The actions of the boson operators on $\ket{k^{r_k}}$ are \cite{merzbacher1998quantum}
\begin{equation}
	a_k^{\dagger} \ket{k^{r_k}} = \sqrt{r_k+1} \ket{k^{r_k+1}} , \ a_k \ket{k^{r_k}} = \sqrt{r_k} \ket{k^{r_k-1}} , \label{eq:creation_and_Annil_operators}
\end{equation}
where we have $a_k \ket{k^{0}} = 0$ with $ r_k =0 $ denoting zero occupation of the mode $k$. 
In this sense, we have the normalized state $\ket{k^{r_k}}$ as
\begin{equation}
	\ket{k^{r_k}} = \frac{1}{\sqrt{r_k!}} \big( a_k^{\dagger} \big)^{r_k} \ket{k^0} . 
\end{equation}

We note that the underbraced term in Eq. \eqref{eq:H_1to1} is logarithmically divergent \cite{Harindranath:1987db,PhysRevD.13.2778}. 
One can remove this logarithmic divergence by considering the normal-ordered Hamiltonian \cite{PhysRevD.11.2088}, which means that one ignores a divergent constant from the definition of the mass parameter in the Hamiltonian.

The $(\phi ^4)_2$ theory has also been used for studying spontaneous symmetry breaking. 
However, spontaneous symmetry breaking arises at a critical coupling ${\lambda}/{m^2}$ whose value depends on the treatment of the zero mode \cite{PhysRevD.94.065006,Anand:2017yij,PhysRevD.105.016020,Chabysheva:2022duu}. Evaluating the precise value for the critical coupling involves taking the continuum limit which is very challenging on classical computers due to the nature of the strong correlation in the quantum many-body problems. Quantum computing might be promising for addressing such challenges.

\subsection{Basis choice and eigenvalue problem}

Following the notation of Ref. \cite{Harindranath:1987db}, we denote a general Fock state as $  \ket{ k^{r_k} , \ l^{r_l}, \ m ^{r_m}, \ \cdots  } $.
The total longitudinal momentum of the Fock state is
$ k r_k + l r_l + m r_m + \cdots $, which receives contributions from every single mode in the Fock state. 

As $K$ commutes with $H$, the LF Hamiltonian matrix elements between the Fock states with different $K$ values vanish.
Therefore, in practical DLCQ calculations, one can enumerate and retain the Fock bases with the fixed total longitudinal momentum \cite{Vary:2009gt,Honkanen:2010rc,Wiecki:2014ola,Zhao:2014xaa,Li:2017mlw,Li:2015zda}. 
Within the Fock bases set $\mathcal{B}_K \equiv \{ \ket{ k^{r_k} , \ l^{r_l}, \ m ^{r_m}, \ \cdots  } \} $ of a specific $K$ value, one evaluates the matrix of $H$, or equivalently, the matrix of $\hat{M}^2$ [Eq. \eqref{eq:m2_op}], and computes the mass spectrum of the $(\phi ^4)_2 $ theory. Based on the mass spectrum, one can interrogate the critical coupling constant for the vanishing mass gap of the many-boson system as a function of $K$, and extrapolate for the critical coupling in the limit of $K\rightarrow \infty $, i.e., the continuum limit \cite{Harindranath:1987db,PhysRevD.105.016020}.

\section{Hamiltonian input scheme}
\label{sec:Ham_Input}

In this section, we present the elements of our Hamiltonian input scheme.
We begin with our strategy to squeeze the boson operators, and rewrite the many-boson Hamiltonian with the squeezed boson operators as well as their combinations.
Then, we discuss the scheme to encode the many-boson states, and present the design of the circuit modules for the squeezed boson operators as well as their combinations.
Based on these modules, we show our Hamiltonian input scheme utilizing the idea of quantum walk.
Finally, we preset the analysis of the qubit and gate cost of our input scheme.

\subsection{Modified Hamiltonian}

For the $(\phi ^4 )_2$ theory, we can rewrite Eq. \eqref{eq:original_Phi4_Hamiltonian} according to the unique types of monomials as
\begin{align}
	H = \underbrace{ B_{j_0} a^{\dagger}_k a_k + \cdots }_{H_{1\rightarrow 1}} + \underbrace{ B_{j_1} a^{\dagger}_k a^{\dagger}_{l} a_m a_n + \cdots }_{H_{2\rightarrow 2}} + \underbrace{ B_{j_2} a^{\dagger} _k a_l a_m a_n + \cdots }_{H_{3\rightarrow 1}} + \underbrace{ B_{j_3} a^{\dagger} _n a^{\dagger} _m a^{\dagger} _l a_k + \cdots }_{H_{1\rightarrow 3}} ,
	\label{eq:original_Hamiltonian}
\end{align}
where we employ the subscripts $k$, $l$, $m$, $n \in [1,\ K]$ to specify the modes of the boson operators. $B_{j_i}$ denotes the coefficient, with
\begin{equation}
	B_{j_0} = \frac{1}{k}, \ B_{j_1} = \frac{\lambda}{4\pi} \frac{1}{N^2_{kl} N^2_{mn}} \frac{\delta _{m+n, k+l}}{\sqrt{klmn}} , \ B_{j_2} = B_{j_3} = \frac{\lambda}{4\pi} \frac{1}{N^2_{lmn}} \frac{\delta _{m+n+l, k}}{\sqrt{klmn}} .
\end{equation}
The dots denote the other monomials with the same types of operator structures as those shown explicitly.

In order to develop our input scheme for the many-boson Hamiltonian, we squeeze (or scale) the boson operators for the mode $k$ according to the maximal occupation value $\Lambda _k = \lfloor K/k \rfloor $ that is consistent with the total longitudinal momentum $K$ for the system as
\begin{equation}
	b_k \equiv {a_k }/{\sqrt{ \Lambda _k }}, \ b^{\dagger} _k = {a^{\dagger}_k}/{\sqrt{\Lambda _k}} . \label{eq:scaled_ladder_Ops}
\end{equation}
This leads to the following identities
\begin{align}
		b_k \ket{k^{r_k} } = \sqrt{ \frac{r_k}{\Lambda _k} } \ket{ k^{r_k-1} }, \ & \ \text{for} \ r \in [1, \Lambda _k] , \label{eq: annil_scaling} \\
		b^{\dagger}_k \ket{k^{r_k} } = \sqrt{ \frac{r_k+1}{ \Lambda _k } } \ket{ k^{r_k+1} } , \ & \ \text{for} \ r \in [0, \Lambda _k -1]  
		\label{eq: creation_scaling},
\end{align}
with $b_k \ket{k^ 0} = 0  $.
According to this scaling, we have the scaled normalization factor $ \sqrt{r_k/\Lambda _k } \in (0, 1] $ for $r_k \in [1, \Lambda _k] $ in Eq. \eqref{eq: annil_scaling}, and  the factor $\sqrt{(r_k+1)/\Lambda _k} \in (0, 1] $ for $ r_k \in [0, \Lambda _k -1] $ in Eq. \eqref{eq: creation_scaling}. These scaled normalization factors, denoted as $\xi _k$, can be encoded in terms of the angles of the rotational gates \cite{Qiskit} in the quantum circuit, as shown in the following text.

With the squeezed boson operators, we can rewrite $H $ [Eq. \eqref{eq:original_Hamiltonian}] as
\begin{equation}
	H = \underbrace{ B'_{j_0} b^{\dagger}_k b_k + \cdots }_{H_{1\rightarrow 1}} + \underbrace{ B'_{j_1} b^{\dagger}_k b^{\dagger}_{l} b_m b_n + \cdots }_{H_{2\rightarrow 2}} + \underbrace{ B'_{j_2} b^{\dagger} _k b_l b_m b_n + \cdots }_{H_{3\rightarrow 1}} + \underbrace{ B'_{j_3} b^{\dagger} _n b^{\dagger} _m b^{\dagger} _l b_k + \cdots }_{H_{1\rightarrow 3}} ,
	\label{eq:modified_H}
\end{equation}
with
\begin{align}
	 B'_{j_0} =& \frac{\Lambda _k} {k}, \\
	 B'_{j_1} =& \frac{\lambda}{4\pi} \frac{1}{N^2_{kl} N^2_{mn}} \sqrt{\frac{\Lambda _k \Lambda _l \Lambda _m \Lambda _n}{klmn} } \delta _{m+n, k+l} , \\
	 B'_{j_2} =& B'_{j_3} = \frac{\lambda}{4\pi} \frac{1}{N^2_{lmn}} \sqrt{\frac{\Lambda _k \Lambda _l \Lambda _m \Lambda _n}{klmn} } \delta _{m+n+l, k} .
\end{align}
The boson modes $k, \ l, \ m, \ n$ in $H$ are not necessarily different. In this sense, we can group the squeezed boson operators according to the modes in each monomial, while preserving the ordering of the creation and annihilation operators in view of the commutation relations. 
For example, with $k \neq l \neq m$, we can rewrite $b ^{\dagger}_k  b^{\dagger}_l b_l b_k $ and $b ^{\dagger}_k  b^{\dagger}_k b_l b_m$ as $ (b ^{\dagger}_k b_k) (b^{\dagger}_l b_l) $ and $(b ^{\dagger}_k  b^{\dagger}_k) (b_l) (b_m)$, respectively.
With this grouping for all the monomials in $H$, we find that there are eight unique combinations $\{W_k \}$ of the squeezed boson operators that act on one single mode. 
We list these unique combinations in Table \ref{tab:all_items}, where we also present other details of these combinations.

\begin{table}[ht]
	\caption{Eight unique combinations of the squeezed boson operators. The first column lists all the unique combinations $W_k$ of the squeezed boson operators. The second column presents the notations $(\cdot )_k$ of the circuit modules for $W_k$. The third column presents the scaled normalization factors $\xi _{r_k}$ of the many-boson state obtained from the action of $W_k$ on $\ket{ k^{r_k} }$ [Eq. \eqref{eq:action_of_Wk}] for $r_k \in [r_{k,i}, r_{k,f}]$ (fourth column). The fifth column presents the $r_k$ values that belong to the vanishing cases in the practical circuit design of $(\cdot)_k$. The rightmost column presents the increment and decrement $\Delta _k$ in the occupation $r_k$ when $W_k$ acts on $\ket{ k^{r_k} }$.}
	\begin{tabular}{lllllc}
		\hline \hline
		$W_k$      \ \ \ \ \ \  & $(\cdot )_k$ \ \ \ \ \ \ \ \ \ \ \ & scaled normalization factors $\xi _{r_k}$ \ \  & $ [r_{k,i}, r_{k,f}] \ \ \ \ \ $ & vanishing cases \ \ \ \ \ \ \ \ \   & $\Delta _k$ \\ \hline
		$b^{\dagger}_k$ & $(+)_k$ & $\sqrt{ (r_k +1)/\Lambda _k }$ &  $[0,\ \Lambda_k -1]$ &  $ \Lambda _k$  & $ { + 1} $ \\
		$b^{\dagger}_k b^{\dagger}_k$ & $(++)_k$ & $\sqrt{ (r_k +1) (r_k +2) /\Lambda _k^2  }$  & $[0,\ \Lambda_k -2]$  & $ \Lambda _k -1, \ \Lambda _k $ & $ {+ 2} $  \\
		$b^{\dagger}_k b^{\dagger}_k b^{\dagger}_k $ &  $(+++)_k$    & $ \sqrt{ (r_k +1) (r_k +2) (r_k +3) /\Lambda _k ^3 }$  & $[0,\ \Lambda_k -3]$ & $ \Lambda _k -2, \ \Lambda _k -1, \ \Lambda _k $  & $ {+ 3}$  \\
		$ b_k $ &  $(-)_k$ & $\sqrt{r_k / \Lambda _k}$  & $ [1,\ \Lambda _k] $ &  $0$  &  $ { - 1}$ \\
		$ b_k b_k $ & $(--)_k$ & $ \sqrt{r_k (r_k -1) /\Lambda ^2 _k } $ & $ [2,\ \Lambda _k] $ & $ 0 ,\ 1$ &  ${-2}$ \\
		$ b_k b_k b_k $  & $(---)_k$ &  $\sqrt{r_k (r_k -1) (r_k -2) /\Lambda ^3 _k } $ & $ [3,\ \Lambda _k] $ & $  0 ,\ 1, \ 2 $ &  $ {- 3}$ \\
		$b^{\dagger}_k b_k$ & $(+-)_k$ & $ r_k/\Lambda _k $ & $ [1,\ \Lambda _k] $ & $  0 $  &  ${  0}$ \\
		$b^{\dagger}_k b^{\dagger}_k b_k b_k$ & $(++--)_k$ & $r_k (r_k -1 ) / \Lambda _k ^2 $ & $ [2,\ \Lambda _k] $ & $ 0, \ 1$ &  ${  0}$ \\
		\hline \hline
	\end{tabular}
	\label{tab:all_items}
\end{table}

\subsection{Basis encoding scheme}
\label{sec:basis_encoding_scheme}

Recall that we retain the Fock basis set $\mathcal{B}_K$ with fixed total longitudinal momentum $K$ and omit the zero mode in this work. In order to encode the Fock states in $ \mathcal{B}_K $, we choose to implement the ``system" register (denoted by ``$s$") that consists of a set of subregisters $\{ s_k \}$: for the mode $k \in [1, K] $, we apply the compact mapping to encode the occupation $r_k$ as binaries in the subregister $s_k$. 
Following this scheme, as an example, we can encode the Fock state $ \ket{\mathcal{F}} = \ket{k^{r_k}, l^{r_l}, m^{r_m } }$ with $1\leq k < l < m \leq K$ as the register state
\begin{equation}
	\ket{0}_{s_1} \cdots \ket{{r_k}}_{s_k} \cdots \ket{{r_l}}_{s_l} \cdots \ket{{r_m}}_{s_m} \cdots \ket{0}_{s_K} ,
\end{equation}
where the occupations $r_k$, $r_l$, and $r_m$ are encoded in the binary representation. We adopt the notation $\ket{\cdot} _{reg}$ to specify the state of the quantum register ``$reg$". We occasionally omit the subscript when there is no risk of confusion.

\subsection{Circuit representation of $W_k$}
\label{sec:module_design}

\begin{figure}[ht] 
	\centering
	\includegraphics[width=0.65\linewidth]{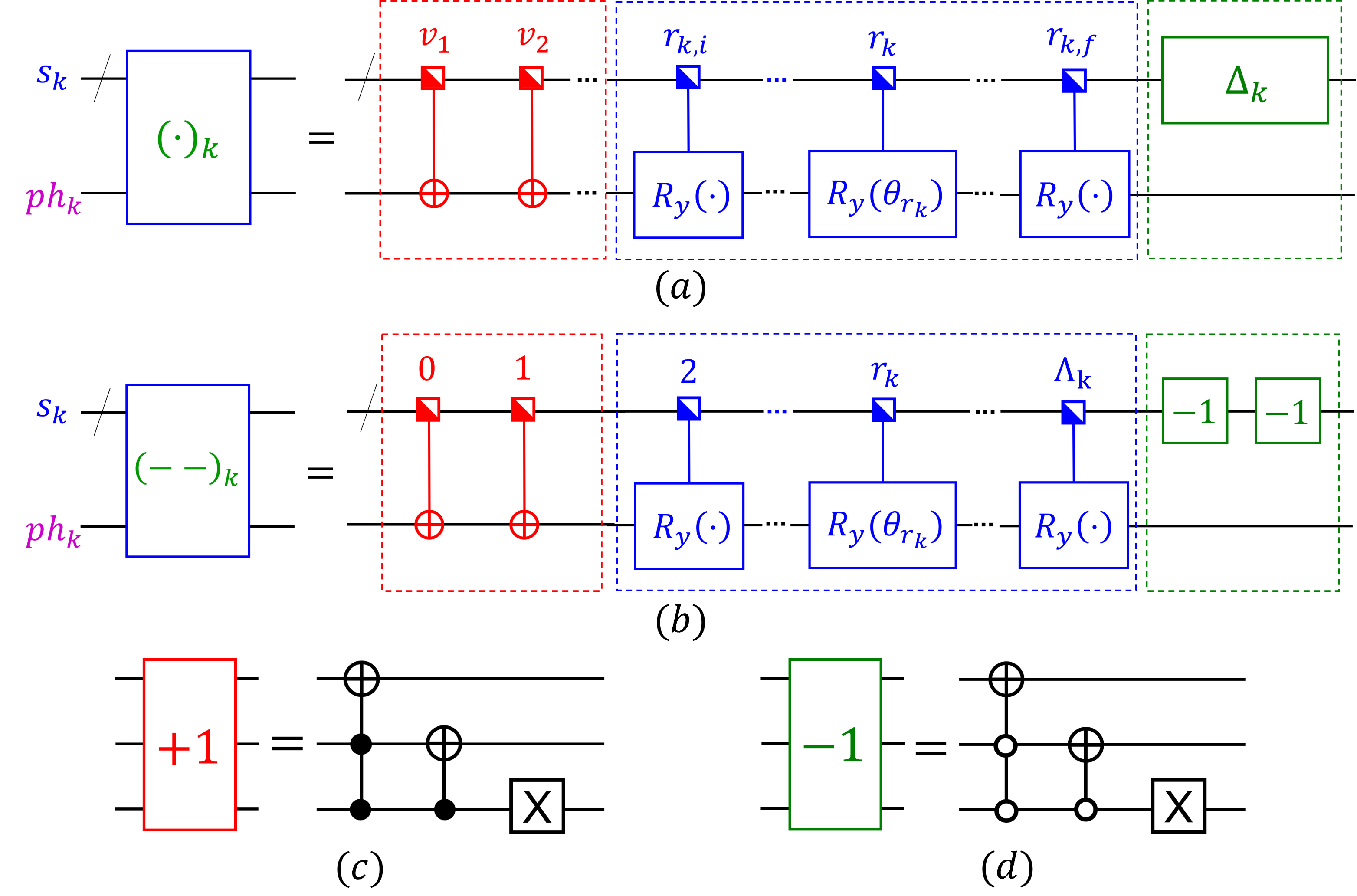}
	\caption{(color online) 
		Schematic design of the circuit module $(\cdot)_k$ of $W_k$ in Table \ref{tab:all_items}. 
		Panel (a) presents the general scheme of the circuit design of $(\cdot)_k$, where $v_1$, $v_2$, $ \cdots$ denote the occupation numbers for the vanishing cases. 
		Panel (b) shows an illustration of the circuit design of the module $(--)_k$ that corresponds to $W_k = b_k b_k$.
		Panels (c) and (d) are the examples of the circuit designs of the quantum adder ``$+1$" and subtractor ``$-1$" for one-unit increment and decrement in the three-qubit case, respectively. 
		The slashed line denotes a register that may contain several qubits.
		The half-filled squares denote the controlling conditions for the quantum operations; these conditions can be constructed according to corresponding occupation values above the squares. 
		The dot in each $R_y $ gate denotes the corresponding rotational angle.
	} 
	\label{fig:operator_circuit}
\end{figure}

We discuss the design of the circuit module $(\cdot)_k$ for the combination $W_k$ of the squeezed boson operators in this section. Indeed, this module can be understood as the ``circuit representation" of $W_k$.
A schematic circuit design of $( \cdot )_k$ is shown in Fig. \ref{fig:operator_circuit}(a).

We recall that the occupation $r_k $ of the mode $k$ is encoded as binaries in the subregister $s_k$. For each $s_k$, we employ a single qubit $ph_k $, which is initialized as $ \ket{0} $.  Here, the qubit $ ph_k $ encodes the scaled normalization factor resulting from the action of $W_k$ on the state $\ket{k^{r_k}}$ of mode $k$ [Table \ref{tab:all_items}].
Jointly, the set of qubits $\{ ph_k \} $ make up the ``phase" register (denoted as ``$ph$").
Then, we design the circuit module $( \cdot )_k$ to conduct the following operations on the register states 
\begin{equation}
	 \ket{{r_k}}_{s_k } \ket{0} _{ph_k}  \xrightarrow{( \cdot )_k } \ket{ {r_k + \Delta _k } }_{s_k} \ket{\gamma _{r_k} } _{ph_k} ,
	\label{eq:W_k_function}
\end{equation}
where $\ket{\gamma _{r_k} } = \ket{\xi _{r_k} } = \xi _{r_k} \ket{0} + \sqrt{1-\xi ^2 _{r_k}} \ket{1 } $ for $r_k \in [r_{k,i}, \ r_{k,f}] $ (fourth column in Table \ref{tab:all_items}). Here, $\xi _{r_k} \in (0,1] $ is the scaled normalization factor and $\Delta _k$ denotes the change in $r_k$. The values of $\xi _{r_k} $ and $\Delta _k$ are shown in the third column and the rightmost column in Table \ref{tab:all_items}, respectively. In practical design of the circuit module $(\cdot)_k$, it is convenient to specify the ``vanishing cases"  (fifth column in Table \ref{tab:all_items}). These correspond to the cases where $ W_k \ket{k^{r_k}} =0 $ or $ W_k \ket{ k^{r_k} } \propto \ket{ k^{r_{k} + \Delta _k} } $ with $ (r_k + \Delta _k) > \Lambda _k = \lfloor K/k \rfloor $.
We set $ \ket{\gamma _{r_k} } = \ket{1} $ for the $r_k$ values that belong to such vanishing cases. 

The circuit module $(\cdot)_k$ of the combination $W_k$ can be achieved by the following analyses and constructions.
\paragraph*{1. Vanishing cases.}
For each $r_k$ value that belongs to the vanishing cases (fifth column in Table \ref{tab:all_items}), we flip the state of $ph_k$ from $\ket{0}$ to $\ket{1}$ controlled by the subregister $s_k$ being in the state $\ket{r_k}_{s_k}$.

\paragraph*{2. Nonvanishing cases.}
For the nonvanishing cases with $r_k \in [r_{k,i},\ r_{k,f}]$ (fourth column in Table \ref{tab:all_items}), we have 
\begin{equation}
	W_k \ket{k^{r_k}} = \xi _{r_k} \ket{ k^{r_k + \Delta _k} } .
	\label{eq:action_of_Wk}
\end{equation}
As for the corresponding circuit construction, we apply a set of $R_y (\theta _{r_k}) $ gates \cite{Qiskit} to the qubit $ph_k$, each of which is controlled by the state $ \ket{ r_k } _{s_k}$ with $r_k \in [r_{k,i},\ r_{k_f}]$. We take the rotational angles as
\begin{equation}
	\theta _{r_k} = 2 \arccos \xi _{r_k} ,
\end{equation}
such that the state of $ph_k$ becomes $ \ket{\xi _{r_k} } _{ph_k} \equiv R_y (\theta _{r_k}) \ket{0} _{ph_k}  $.
That is, $\xi _{r_k} $ is encoded as the coefficient of the component $\ket{0} $ of the state $ \ket{\xi _{r_k} } _{ph_k} $.

\paragraph*{3. Changes in the occupation.}
To account for the change $ \Delta _k $ in the occupation $r_k$ recorded as the state $\ket{r_k}_{s_k}$, we can perform a sequence of quantum adders or subtractors \cite{rieffel2011quantum} to the subregister $s_k$. These quantum arithmetic operators perform the following operation on the state of $s_k$:
\begin{equation}
	\ket{r_k} _{s_k} \rightarrow \ket{r_k + \Delta _k} _{s_k} .
\end{equation}
We illustrate the circuits of the quantum adder and subtractor in Figs. \ref{fig:operator_circuit}(c) and (d) for the three-qubit case, respectively. 

The above completes the construction of the circuit module $(\cdot)_k $ corresponding to Eq. \eqref{eq:W_k_function}.
As a specific example, we present in Fig. \ref{fig:operator_circuit}(b) the circuit module $(--)_k$ of $W_k = b_k b_k$.

\subsection{Encoding many-boson Hamiltonian via quantum walk}


We now present the input scheme of the LF Hamiltonian of the $ (\phi ^4)_2 $ theory, utilizing the concept of the quantum walk \cite{PhysRevLett.102.180501,childs2010relationship,berry2012black}. We show our construction of the quantum walk states [Eqs. \eqref{eq:forward_walk_state} and \eqref{eq:back_walk_state}] according to the information of the Hamiltonian. We also prove that these so-constructed quantum walk states correctly encode the Hamiltonian [Eqs. \eqref{eq:inner_product} and \eqref{eq:Hamiltonian_input_scheme}].

We first construct the {\it forward} walk state. Provided the Fock state $\ket{\mathcal{F}}$, we start with the state of multiple quantum registers
\begin{equation}
	\ket{\Phi _0} = \ket{0} _{id} \ket{ \mathcal{F} } _{s} \ket{0} _{ph} \ket{0}_{ me}  \ket{1} _{ac} ,
	\label{eq:initialized_WalkState}
\end{equation}
where the index register ``$id$" is initialized as $\ket{0}$. The Fock state $ \ket{\mathcal{F}} $ is encoded by the system register $s$ that consists of a set of subregisters $ \{s_k \} $ (with $k \in [1, K]$).
The phase register $ph$ contains $K$ qubits $\{ ph _k \}$.
Each single-qubit subregister $ph_k$ is initialized as $\ket{0}$ and is coupled with the subregister $s_k$. 
In addition, we adopt the single-qubit ``matrix-element" register (denoted by ``$me$") that is initialized as $\ket{0}$ to encode the coefficients $\{B'_j \}$ of the monomials of $H$ [Eq. \eqref{eq:modified_H}]. Finally, we employ the single-qubit ``action" register (denoted as ``$ac$") and initialize it as $\ket{1}$. The purpose of introducing the register $ac$ will become clear in the following text.

Based on $\ket{\Phi _0}$, we introduce the diffusion operator (which consists of a set of Hadamard gates) to the register $id$ and obtain 
\begin{equation}
	\ket{\Phi _1} =\frac{1}{\sqrt{\mathcal{D}}} \sum _{j=0} ^{\mathcal{D}-1} \ket{j} _{ id} \ket{ \mathcal{F} } _{s} \ket{0} _{ph} \ket{0}_{me} \ket{1} _{ac} .
\end{equation}

We label each monomial in $H$ [Eq. \eqref{eq:modified_H}] by a specific index $j \in [0, \mathcal{D}-1] $. For the clarity of explanation, we take the $j$th monomial of $H$ [Eq. \eqref{eq:modified_H}] to be $B'_j W_k W_l $ (with $1\leq k<l \leq K$) as an example. Then, we implement the following operations controlled by $\ket{j}_{id}$. 
\paragraph*{Step 1.}
We apply the circuit module $(\cdot )_k$ of $W_k$ to the subregisters $s_k$ and $ph_k$, and the circuit module $(\cdot )_l$ of $W_l$ to $s_l$ and $ph_l$. 
According to Eq. \eqref{eq:W_k_function}, we have the operations on the state of registers $s$ and $ph$ as
\begin{equation}
	 \ket{\mathcal{F}}_s \ket{0}_{ph} \xrightarrow{(\cdot)_l (\cdot)_k}  \ket{\mathcal{F}_j}_s \ket{\Upsilon _j}_{ph} .
\end{equation}
Here, the Fock state $\ket{\mathcal{F}}$ is encoded as $ \ket{r_1}_{s_1} \cdots \ket{r_k}_{s_k} \cdots \ket{r_l}_{s_l} \cdots \ket{r_K}_{s_K} $ in the register $s$. With the changes of the occupations in the modes $k$ and $l$ determined by $W_k$ and $W_l$, we have the modified Fock state  $ \ket{\mathcal{F}_j} $ encoded as $ \ket{r_1}_{s_1} \cdots \ket{r_k + \Delta _k}_{s_k} \cdots \ket{r_l + \Delta _l }_{s_l} \cdots \ket{r_K}_{s_K} $ in the register $s$. We also have the operation on the phase register $ph$ as
\begin{align}
	 \ket{0}_{ph} & = \ket{0}_{ph_1} \cdots \ket{ 0 }_{ph_k} \cdots \ket{ 0 }_{ph_l} \cdots \ket{0} _{ph_K} \nonumber \\
	\rightarrow  \ket{\Upsilon _j}_{ph} & \equiv \ket{0}_{ph_1} \cdots \ket{ \gamma _{r_k} }_{ph_k} \cdots \ket{ \gamma _{r_l} }_{ph_l} \cdots \ket{0} _{ph_K} .
\end{align}

\paragraph*{Step 2.}

We operate on the single-qubit register $me$ as
\begin{equation}
	\ket{0}_{me} \rightarrow e^{i \beta _j} \ket{\rho _j} _{me},
		\label{eq:encoding_me}
\end{equation}
where $\ket{\rho _j} \equiv \rho _j \ket{0} +\sqrt{1-\rho ^2_j } \ket{1} $. The parameters $\beta _j $ and $\rho _j $ are determined from the identity
\begin{equation}
	\rho _j e^{i \beta _j} =  B_j' / \Xi ,
\end{equation}
with  $\Xi \geq \max _j |B_j'|$ and $\rho _j = | B_j' / \Xi  | \leq 1$. We take $\beta_j =0 $ as $B_j' > 0$ in the $(\phi ^4)_2$ theory discussed in this work.\footnote{In general applications, we take $\beta _j = \arg [B_j' / \Xi ] \in (-\pi , \pi] $ as $B_j'$ can be a complex number.}
In other words, the coefficient $B_j' $, scaled by $\Xi$ that is at least the max norm of all the coefficients $ \{ B'_j\} $ of $H$ [Eq. \eqref{eq:modified_H}], is encoded by the register $me$ as the coefficient of the component $\ket{0}$ of the state $e^{i \beta _j} \ket{\rho _j} _{me}$. 
Practically, the operation in Eq. \eqref{eq:encoding_me} can be achieved by applying the phase gate $P_X(\beta _j)$ followed by a $R_y (\alpha _j)$ gate to the register $me$, where we take 
\begin{equation}
	P_X(\beta _j) \ket{0} = e^{i\beta _j} \ket{0}, \ P_X(\beta _j) \ket{1} = \ket{1}, \ \alpha _j = 2\arccos \rho _j .
\end{equation} 

\paragraph*{Step 3.}
We flip $\ket{1} _{ac} $ to $\ket{0} _{ac}$ by a Pauli $X$ gate. 

The above completes the operations on $\ket{\Phi _1}$, where these operations are based on the assumed form of the $j$th monomial $B'_j W_k W_l $. 

\begin{figure}[ht] 
	\centering
	\includegraphics[width=0.50\linewidth]{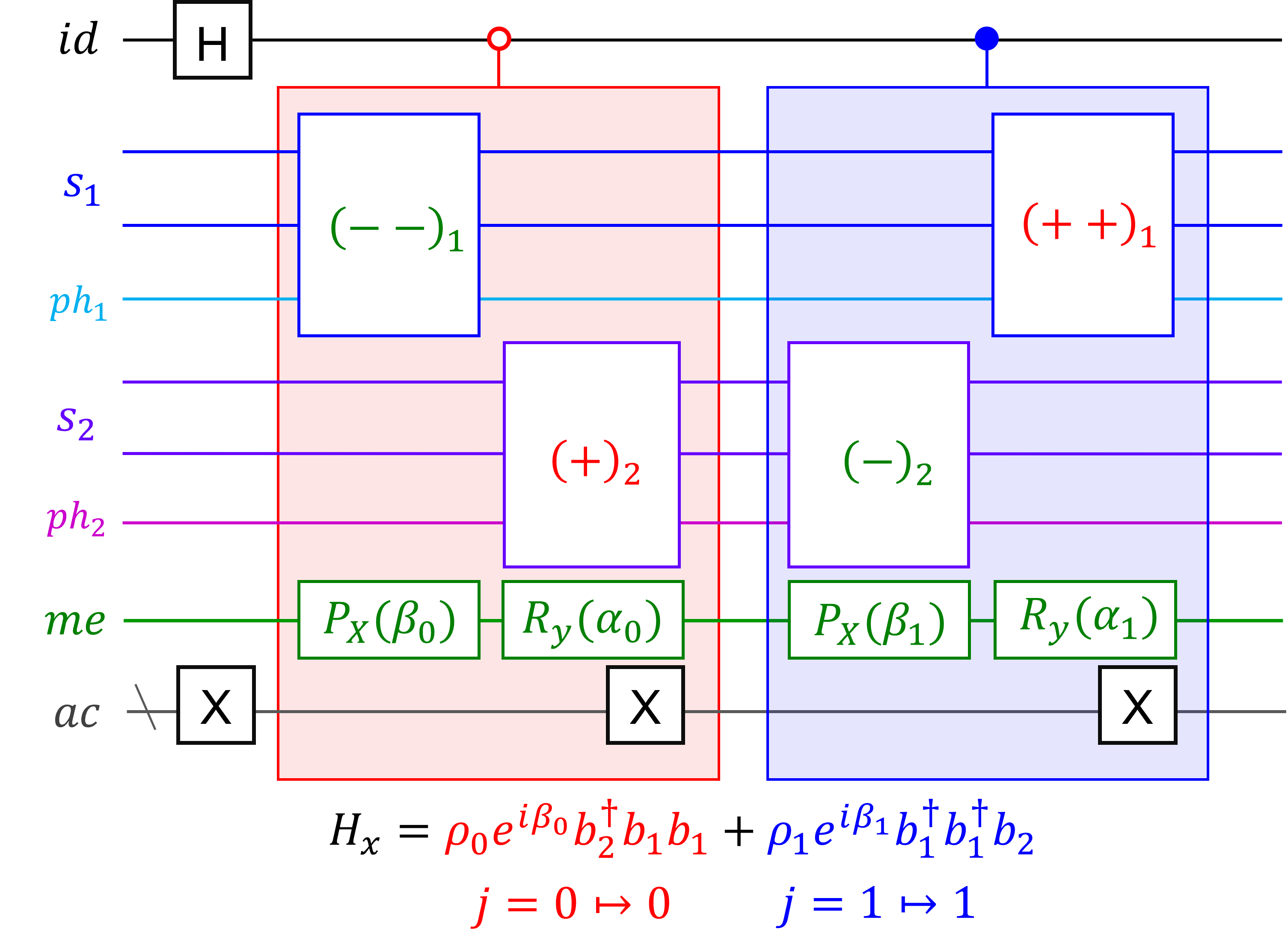}
	\caption{(color online) Illustration of the circuit design of the forward walk state according to the Hamiltonian $H_x$ of a simple many-boson system. The combinations of the squeezed boson operators and their circuit modules are shown in Table \ref{tab:all_items} and Fig. \ref{fig:operator_circuit}(a), respectively. The coefficient in the $j$th monomial (with $j=0,1$) is expressed as $\rho _j e^{i\beta _j}$ with $\rho _j \in (0,1]$; it is encoded by a phase gate $P_X(\beta _j) $ with $P_X(\beta _j) \ket{0} = e ^{i \beta _j } \ket{0} $ and $P_X(\beta _j) \ket{1} = \ket{1}$. The angle of the $R_y$ gate \cite{Qiskit} is $\alpha _j = 2\arccos \rho _j$.
	For clarity, the subregisters $s_k$ and $ph_k$ with $k=1,2$ are coupled in the circuit. 
	With the single-qubit register $id$ in this example, the hollow red circle and the filled blue circle denote the operations controlled by $j=0$ and $j=1$, respectively. 
	} 
	\label{fig:example_two_terms}
\end{figure}

Analogously, we apply the same steps as described above for all the monomials in $H$ [Eq. \eqref{eq:modified_H}], where each of these monomials is labeled by a distinct $j$. 
With these operations, we can construct the {\it forward} walk state
\begin{equation}
	\ket{\Phi _2} = \frac{1}{\sqrt{\mathcal{D}}} \sum _{j=0} ^{\mathcal{D}-1} \ket{j} _{ id} \ket{ \mathcal{F}_j } _{s} \ket{ \Upsilon _j } _{ph} e^{i \beta _j} \ket{\rho _j} _{ me} \ket{ b_j } _{ac} .
	\label{eq:forward_walk_state}
\end{equation}
Here we have $b_j = 0$ if the index $j$ indeed corresponds to a monomial in the Hamiltonian $H$; $b_j$ remains as $1$ otherwise. In this sense, the action register $ac$ facilitates tracking the null correlations to the monomials in $H$ as some of the $j$ values encoded in the register $id$ may not relate to any monomial in $H$ in practical applications.
In Fig. \ref{fig:example_two_terms}, we illustrate the circuit to construct the forward walk state via a simple example.

Next, we construct the {\it backward} walk state based on the Fock state $\ket{\mathcal{G}}$ as 
\begin{equation}
	\ket{\Psi _0}  = \ket{0} _{id} \ket{ \mathcal{G} } _{s} \ket{0} _{ ph} \ket{0}_{ me} \ket{0} _{ ac} ,
\end{equation}
where the register $ac$ is initialized as $\ket{0}$ in this case.
With the action of the diffusion operator on $ \ket{0} _{id} $, we get
\begin{equation}
	\ket{\Psi _1} = \frac{1}{\sqrt{\mathcal{D}}} \sum _{k=0} ^{\mathcal{D}-1} \ket{k} _{ id} \ket{ \mathcal{G} } _{s} \ket{0} _{ph} \ket{0}_{me} \ket{0} _{ac} .
	\label{eq:back_walk_state}
\end{equation}

The walk states $\ket{\Phi _2} $ and $\ket{\Psi _1} $ block encode the many-boson Hamiltonian. In particular, the inner product of the walk states is
\begin{align}
	\langle \Psi_1 | \Phi_2 \rangle =&  \frac{1}{\mathcal{D}} \sum _{j=0} ^{\mathcal{D}-1} \sum _{k=0} ^{\mathcal{D}-1} e^{i \beta _j} \langle k|j \rangle _{ id} \langle \mathcal{G} | \mathcal{F}_j \rangle _{s} \langle 0| \Upsilon _j \rangle _{ph}  \langle 0 | \rho _j \rangle _{me}  \langle 0 |b_j \rangle _{ac} ,
	\label{eq:inner_product}
\end{align}
which correctly produces the many-boson matrix element of the scaled Hamiltonian $H' \equiv H/(\mathcal{D}\Xi)$ based on the Fock states $\ket{\mathcal{F}}$ and $\ket{\mathcal{G}}$.
The kernel $  \langle 0 |b_j \rangle _{ac} = \delta _{b_j,0} $ removes the null contributions indexed by the $j$ values that do not correspond to any monomial in the LF Hamiltonian.

We can rewrite the walk states based on the isometries $\mathcal{T}_1$ and $\mathcal{T}_2$ as
\begin{equation}
	\ket{\Phi _2} = \mathcal{T}_1 \ket{\mathcal{F}}_s \otimes \ket{0} _a , \ \ket{\Psi _1 } = \mathcal{T} _2 \ket{\mathcal{G}}_s \otimes \ket{0} _a ,
\end{equation}
where $a$ denotes all the registers except the system register $s$.
These walk states block encode the many-boson Hamiltonian as
\begin{equation}
	( \bra{\mathcal{G}} _s \otimes \bra{0} _a ) \mathcal{T} ^{\dagger} _2 \mathcal{T}_1 (\ket{\mathcal{F}}_s \otimes \ket{0} _a ) =  \frac{1}{\mathcal{D} \Xi} \langle \mathcal{G} | H | \mathcal{F} \rangle .
	\label{eq:Hamiltonian_input_scheme}
\end{equation}
We comment that this Hamiltonian input scheme is non-Hermitian, i.e., $\mathcal{T} ^{\dagger} _2 \mathcal{T}_1 \neq \big( \mathcal{T} ^{\dagger} _2 \mathcal{T}_1 \big)^{\dagger}$. 
Compared to the typical sparse-matrix input scheme discussed in Refs. \cite{PhysRevLett.102.180501,childs2010relationship,berry2012black} that access the Hamiltonian matrix elements according to their row and column indices, our input scheme dynamically computes the matrix element of the scaled many-boson Hamiltonian $H'$ based on the Fock states [Eq. \eqref{eq:inner_product}]. 
Finally, we remark that our input scheme does not interact with any precomputed database. There is no need to uncompute any ancilla qubits, where such uncomputation procedures are necessary in the typical Hamiltonian input scheme \cite{PhysRevLett.102.180501,childs2010relationship,berry2012black}.

\subsection{Analysis of the gate and qubit cost}
\label{sec:cost}

We estimate the gate cost of our Hamiltonian input scheme with the following analysis.
We retain $K$ modes in our calculations by setting the total longitudinal momentum to be $K$. In the most complicated case, there are $O(K^4)$ monomials in $H$ [Eq. \eqref{eq:modified_H}].\footnote{We adopt the conventional notations in computer science for the complexity analysis in this work. For the functions $w(x)$ and $v(x)$, $w(x) \in \Theta (v(x))$ denotes $\lim _{x\rightarrow \infty} \frac{w(x)}{v(x)} = c$ with $c$ being some constant. 
The notation $w(x)\in O(v(x))$ means $\lim _{x\rightarrow \infty} \frac{v(x)}{w(x)} = 0$.
} Expressed in terms of $W_k$, each monomial may contain four combinations of the squeezed boson operators (i.e., in the form of $ W_k W_l W_m W_n$). It takes at most $ \Theta (K) $ gates to construct the circuit module for each combination according to our design (Sec. \ref{sec:module_design}). These are the dominant gate costs required to construct the forward walk state. In comparison, the gate cost to construct the backward walk state is much less than the gate cost required for constructing the forward walk state. Hence, the total gate cost to input the many-boson Hamiltonian scales as $\widetilde{O}(K^5)$, where the tilde notation denotes that we suppress the logarithmic factor from compiling the multiple controlled gates \cite{peskin2018introduction,javadiabhari2015scaffcc,10.1145/2597917.2597939}.
We expect that this gate complexity can be significantly reduced in realistic implementations as (1) the number of monomials is much less than $O(K^4)$ due to the symmetries of the system (e.g., the momentum conservation); and (2) the maximal occupation $\Lambda_k = \lfloor K/k \rfloor $ of the mode $k$ can be much less than $K$ when $k$ is large, which reduces the gate cost to construct the circuit module $(\cdot )_k$.

The qubit cost to input the Hamiltonian is dominated by the number of qubits necessary for the registers $s= \{ s_k \}$ and $ph = \{ ph_k \}$ for $ k\in [1,K] $. According to our encoding scheme, we need $ \log _2 ( \Lambda _k +1 ) $ qubits to encode all the possible occupation values in the subregister $s_k$. As the maximal occupation of the mode $k$ is $\Lambda _k = \lfloor K/k  \rfloor $, the number of qubits required for the register $s$ is
\begin{equation}
	Q_s  = \sum _{k =1} ^{K} \lceil \log _2 ( \Lambda _k +1 ) \rceil  = \sum _{k =1} ^{K} \lceil \log _2 ( \lfloor K/k  \rfloor + 1 ) \rceil  .
\end{equation}
We note that $Q_s$ grows approximately as $2K$ for large $K$. In addition, it takes $K$ qubits for the register $ph$.
In sum, the total number of qubits to block encode the many-boson Hamiltonian is approximately $\Theta(K)$, where we omit the minor qubit cost: (1) the two qubits required for the registers $me$ and $ac$; and (2) the $O(4\log _2 K)$ qubits required for the register $id$ to encode all the possible indices $j$ of the monomials in $H$ [Eq. \eqref{eq:modified_H}]. 

Kirby et al. \cite{PhysRevA.104.042607} proposed a general-purpose scheme to input many-body Hamiltonians, where the authors employ the compact basis encoding and implement abstract log-local operations (for, e.g., ordering, squares, sums, products, square roots, and inverse etc.) in their scheme. 
As for encoding the LF Hamiltonian of the $(\phi ^4)_2$ theory, Ref. \cite{PhysRevA.104.042607} reports the cost in the log-local operations to be $O(K^3)$ and the qubit cost to be $O(K\log K)$. 
However, while being hardware specific, compiling these log-local operations into primitive gates is still an open question which requires extensive studies \cite{PhysRevA.104.042607}.
In comparison, we propose the circuit design for our input scheme with straightforward applications of the elementary gates and their corresponding controlled versions; this enables straightforward implementations of our input scheme in near-term quantum hardwares. 
Due to the complexity of achieving the log-local operations employed in Ref. \cite{PhysRevA.104.042607} on quantum computers, it is hard to present a direct comparison between our input scheme and the one in Ref. \cite{PhysRevA.104.042607}. 

Our scheme inputs the second-quantized Hamiltonian of the many-boson system by dynamically computing the many-body matrix elements on quantum computers based on our encoding scheme of the many-boson state (Sec. \ref{sec:basis_encoding_scheme}) and the circuit representation of the combination $W_k$ of the squeezed boson operators (Sec. \ref{sec:module_design}). We expect our circuit representation of $W_k$ can also be adopted in other Hamiltonian input schemes such as the linear combination of unitaries \cite{Childs2012HamiltonianSU,RyanBabbushNJP2016}. This provides insights in developing alternative input schemes of many-boson Hamiltonians in the future.

\section{Hybrid SA-QKSD algorithm for the spectral calculations}
\label{sec:App_SA_QKSD}

\begin{figure}[ht] 
	\centering
	\includegraphics[width=0.6\linewidth]{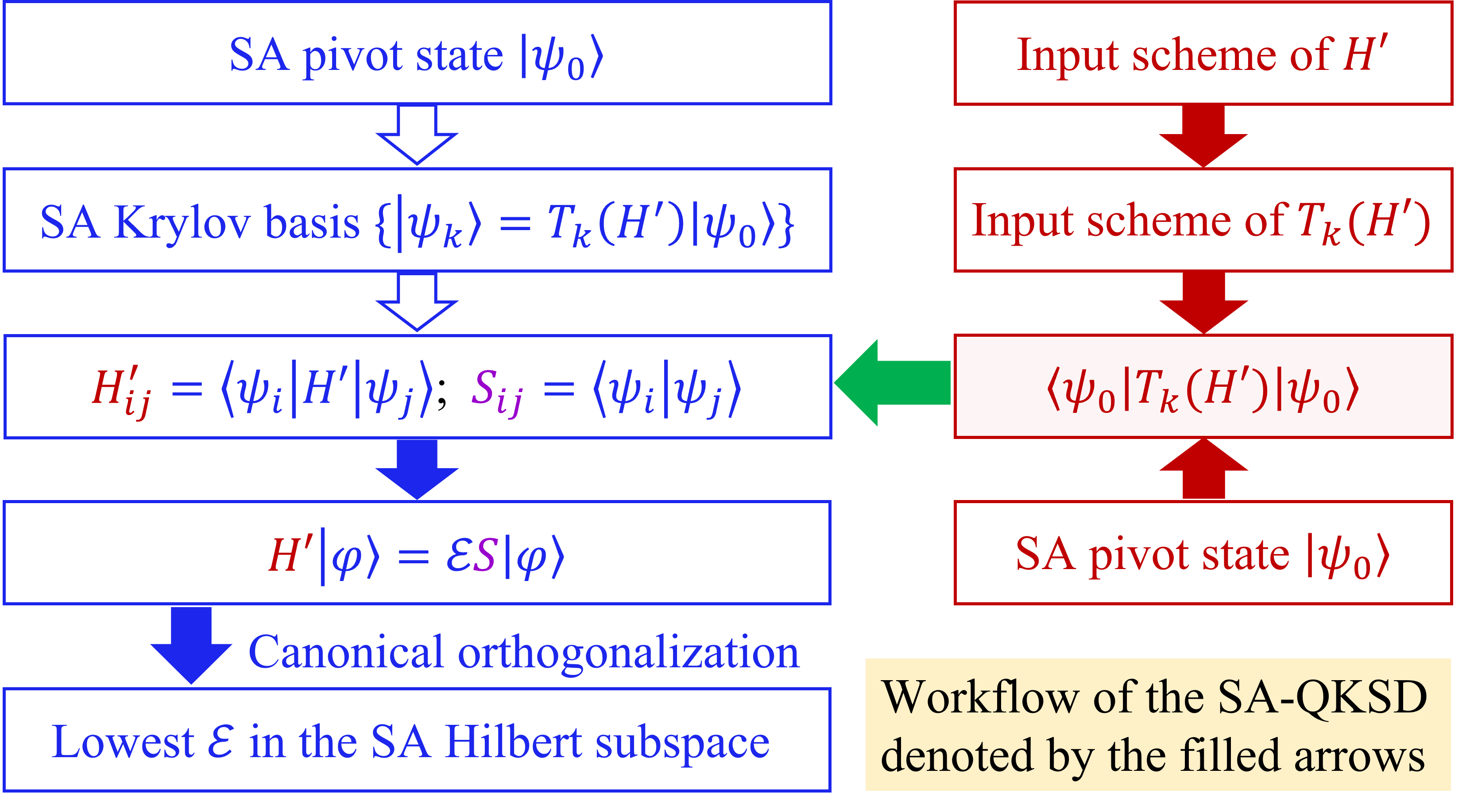}
	\caption{(color online) Workflow of the hybrid quantum-classical SA-QKSD algorithm. The hybrid SA-QKSD approach takes the expectation values [Eq. \eqref{eq:expectationVals}] solved via quantum computing as input to the classical computer, on which one constructs and solves the generalized eigenvalue equation in the Hilbert subspace spanned by the SA Krylov basis set. See text for details. 
	} 
	\label{fig:workflow_hybrid}
\end{figure}

We illustrate the utility of our input scheme for many-boson Hamiltonian [Eq. \eqref{eq:Hamiltonian_input_scheme}] by the spectral calculations of the many-boson system. 
As in this work, we focus implementing the SA-QKSD algorithm \cite{Du:2024zvr,Kirby_2023} for the spectral calculations based on our Hamiltonian input scheme. 
As the workflow illustrated in Fig. \ref{fig:workflow_hybrid}, the SA-QKSD algorithm makes use of the quantum computer to deal with the quantum many-body problems that can be intractable on classical computers, and cast the many-body problems onto a Hilbert subspace spanned by a limited set of Krylov bases that can be handled on the classical computer. Then, one solves the Hamiltonian eigenvalue problem in the Krylov subspace for the eigenenergies of the low-lying states on the classical computer.
Interested readers are also referred to Ref. \cite{motta2023subspace} and references therein for a general review of various other quantum-subspace methods for structure calculations. 
Though we restrict our discussions to the spectral calculations, we comment that our Hamiltonian input scheme can also be implemented with other efficient algorithms, such as the quantum signal processing \cite{PhysRevLett.118.010501}, the quantum singular value transformation \cite{gilyen2019quantum,PRXQuantum.2.040203}, and the quantum eigenvalue transformation \cite{PRXQuantum.3.040305}, etc., for dynamics simulations with optimal gate complexities with respect to the simulation error and time.

\begin{figure}[ht] 
	\centering
	\includegraphics[width=0.40\linewidth]{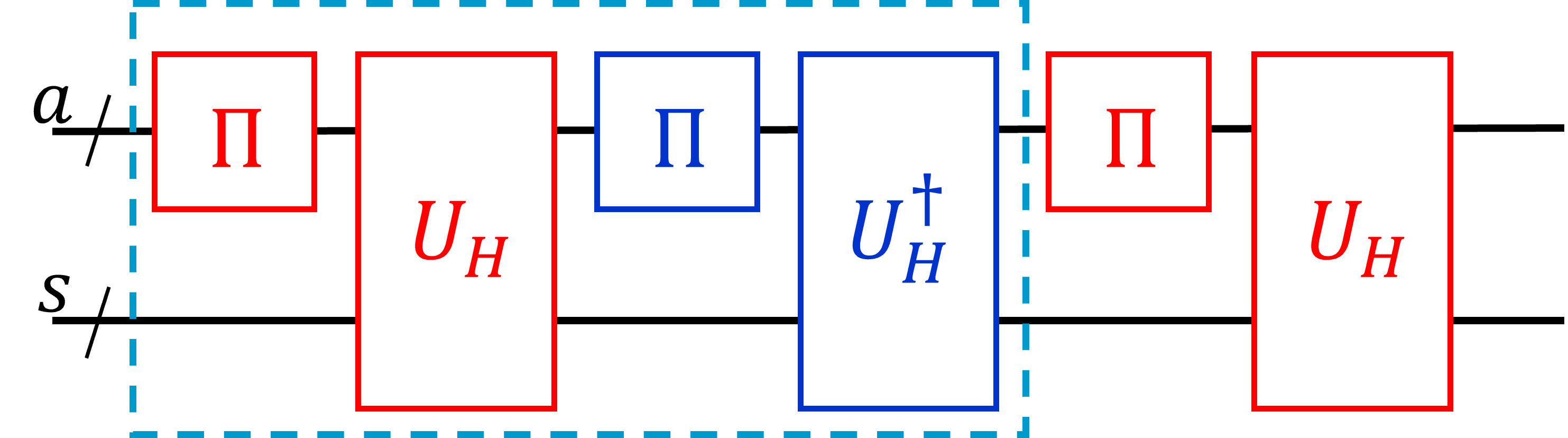}
	\caption{(color online) Circuit to block encode  $ T_3 (H') $. Here $a$ denotes the ancilla register that is initialized as $\ket{0}_a$, while $s$ denotes the system register that encodes the many-body state. The circuit that is enclosed by the dashed blue line block encodes $ T_2 (H') $. 
	} 
	\label{fig:blockEncoding_Chyb}
\end{figure}

With the many-boson Hamiltonian input scheme [Eq. \eqref{eq:Hamiltonian_input_scheme}], we can construct the input scheme of the Chebyshev polynomials of the first kind \cite{kenken2013mathematical} as
\begin{align}
	\langle \mathcal{G} | T_{2k+1}(H') | \mathcal{F} \rangle  
	=& 	( \langle {\mathcal{G} } |_s \otimes \langle 0| _a ) \big[  U_H {\Pi} ( U^{\dagger}_H {\Pi} U_H {\Pi})^k \big] ( | \mathcal{F}  \rangle _s \otimes \ket{0} _a ) \label{eq:Chebyshev_odd} , \\
	\langle \mathcal{G}| T_{2k}(H')  | \mathcal{F} \rangle  
	=&( \langle \mathcal{G} |_s \otimes \langle  0 | _a )  ( U^{\dagger}_H {\Pi} U_H {\Pi})^k  ( | \mathcal{F} \rangle _s \otimes \ket{ 0} _a ) ,
	\label{eq:Chebyshev_even}
\end{align}
with $k=0, \ 1, \ 2, \cdots$. Here we define $ U_H \equiv \mathcal{T} ^{\dagger} _2 \mathcal{T}_1 $. The reflection operator $ 	{\Pi} \equiv (2 |0 \rangle _a \langle 0 | _a - \mathbb{I} _a ) \otimes \mathbb{I} _s $ produces the reflection around $ \ket{0} _a $ in the auxiliary space. The quantum circuit for inputting $T_j(H')$ is achieved via alternative applications of $U_H\Pi$ and $U_H^{\dagger}\Pi$ \cite{lin2022lecture,Du:2024zvr}, where Fig. \ref{fig:blockEncoding_Chyb} presents simple examples for inputting the Chebyshev polynomials.

Utilizing the input scheme of the Chebyshev polynomials, we can construct the Krylov basis as \cite{Kirby_2023}
\begin{equation}
	\{ \ket{\psi _i } =T_i (H') \ket{\psi _0} | i = 0, \ 1, \ 2, \ \cdots, \ \mathcal{K}-1  \} ,
	\label{eq:basis_set_Krylov}
\end{equation}
where $\ket{\psi _0} $ denotes the pivot many-boson state. 

Within the SA-QKSD approach, we can prepare $\ket{\psi _0} $ to be of specific symmetries of the many-boson Hamiltonian --- this can be achieved in a straightforward manner as our input scheme works in the second-quantized representation and the $\ket{\psi _0}$ is encoded in terms of occupations of modes. 
Based on the SA pivot state $\ket{\psi _0}$, we can generate the SA Hilbert subspace that is spanned by the SA Krylov basis $\{ \ket{\psi _i} \} $. Then, we project the Hamiltonian eigenvalue problem to the SA Hilbert subspace and solve for the eigenenergy of corresponding ground state with the elected symmetries, where the unwanted states that do not respect the elected symmetries are naturally excluded.

In particular, the matrix element of $H'$ in the SA Krylov basis is
\begin{equation}
	H'_{ij} = \bra{\psi_i} H' \ket{\psi _j} = \bra{\psi _0 } T_i(H') H' T_j(H') \ket{\psi _0} ,
\end{equation}
while the element of the overlap matrix $S$ is
\begin{equation}
	S _{ij} = \langle \psi _i | \psi _j \rangle =  \bra{\psi _0 } T_i(H') T_j(H') \ket{\psi _0} .
\end{equation}
Utilizing the identities of the Chebyshev polynomials \cite{kenken2013mathematical}
\begin{equation}
	T_0(x) =1 , \ T_1(x) =x ,\ T_i(x) T_j(x) = \big[ T_{i+j} (x) + T_{|i-j|}(x) \big]/2 ,
\end{equation}
one can express $H'_{ij}$ and $S _{ij} $ in terms of the expectation 
\begin{equation}
	\langle T_k (H') \rangle _0 \equiv \bra{\psi _0 } T_k (H') \ket{\psi _0} .
	\label{eq:expectationVals}
\end{equation}

After simplifications, the elements $ H'_{ij} $ and $ S _{ij} $ can be expressed in terms of the expectations $ \langle T_k(H') \rangle _0 \equiv  \bra{\psi _0} T_k(H') \ket{\psi_0} $ as
\begin{align}
	H'_{ij} =& \frac{1}{4} \big[ \langle T_{i+j+1}(H') \rangle _0 + \langle T_{|i+j-1|}(H') \rangle _0 + \langle T_{|i-j+1|}(H') \rangle _0  + \langle T_{|i-j-1|}(H') \rangle _0 \big] , \label{eq:H_ij} \\
	S _{ij} =&  \frac{1}{ 2 } \big[ \langle T_{i+j}(H') \rangle _0 + \langle T_{|i-j|}(H') \rangle _0 \big] .
	\label{eq:S_ij}
\end{align}
The expectation values $\{ \langle T_k (H') \rangle _0 \}$ are, in general, challenging to calculate on classical computers for quantum many-body systems, while they are expected to be evaluated with efficiency on quantum computers \cite{feynman1982simulating} via standard methods such as the Hadamard test \cite{nielsen2010quantum}. 

In the SA-QKSD approach, we quantum compute the expectation values $\{ \langle T_k (H') \rangle _0 \}$. These are then input to the classical computer to solve $\{H'_{ij} \}$ and $\{S _{ij} \} $, based on which we construct the generalized eigenvalue equation 
\begin{equation}
	H' \ket{\varphi } = \mathcal{E} S \ket{\varphi}, 
	\label{eq:GEVP}
\end{equation}
and solve for the eigenvalue $\mathcal{E}$ of the eigenstate $\ket{\varphi}$ in the SA Hilbert subspace that is spanned by $\{ \ket{\psi _i} \} $.

A general bound of $\mathcal{K}$ [Eq. \eqref{eq:basis_set_Krylov}] to compute the ground state energy with the error $\epsilon$ is \cite{Kirby_2023}
\begin{equation}
	\mathcal{K} = \Theta [  ( \log | \kappa _0 |^{-1}  + \log \epsilon ^{-1}  ) \cdot \min ( \epsilon ^{-1}, {\delta}^{-1} )  ] ,
	\label{eq:dim_or_KrylovBasisSpace}
\end{equation}
where $\kappa _0 $ denotes the overlap between the pivot $\ket{\psi _0}$ and the ground state. $\delta $ is the spectral gap between the ground and first-excited state. In practical calculations, one expects a significant reduction in the dimension of the SA Hilbert subspaces for eigenvalues with limited numerical precision. In such cases, it is feasible for the classical computer to evaluate the corresponding generalized eigenvalue equations [Eq. \eqref{eq:GEVP}] without the exposure to the full quantum many-body problem.
Meanwhile, the overlap matrix is ill conditioned in general. In practice, one can adopt the canonical orthogonalization \cite{LOWDIN1970} to approximate the lowest eigenvalue of $H'$ \cite{doi:10.1137/21M145954X} in the SA Hilbert subspace. 

Finally, we estimate the gate and qubit cost for the necessary quantum computation within the hybrid quantum-classical SA-QKSD framework. In particular, it is straightforward to evaluate the expectations values [Eq. \eqref{eq:expectationVals}] via the standard Hadamard test \cite{nielsen2010quantum}. In this sense, the qubit cost for all the circuit is $(Q_s + 1)$. As for the most complex circuit for evaluating the expectation $ \langle T_{ \mathcal{K} } (H') \rangle _0 $, the gate cost scales as $\widetilde{O}(\mathcal{K} \cdot K^5)$.

\section{Numerical demonstration}
\label{sec:example}
Based on our Hamiltonian input scheme, we present the spectral calculations of the LF Hamiltonian of the $(\phi^4)_2$ theory implementing the hybrid quantum-classical SA-QKSD algorithm discussed in Sec. \ref{sec:App_SA_QKSD}.

\subsection{Parameter setting}

We retain a limited model problem with $K=4$ in our numerical demonstration. We elect $m^2=1$ MeV$^2$ and $\lambda /m^2 = 92.4746 $ for the $(\phi ^4)_2$ theory, where the reasons for this parameter setup will be clear in the following text. The LF Hamiltonian for the $(\phi^4)_2$ theory in terms of the squeezed boson operators [Eq. \eqref{eq:scaled_ladder_Ops}] reads 
\begin{align}
	H_{K=4} =&  B'_0 b^{\dagger}_1 b_1 + B'_1 b^{\dagger}_2 b_2 + B'_2 b^{\dagger}_3 b_3 + B'_3 b^{\dagger}_4 b_4 
	+ B'_4 b^{\dagger}_1 b^{\dagger}_1 b_1 b_1 
	+ B'_5 b^{\dagger}_1 b_1 b^{\dagger}_2 b_2 \nonumber \\
	& + B'_6 b^{\dagger}_1 b_1 b^{\dagger}_3 b_3 + B'_7 b_1 b^{\dagger}_2 b^{\dagger}_2 b_3 + B'_8 b^{\dagger}_1 b_2 b_2 b^{\dagger}_3 + B'_9 b^{\dagger}_2 b^{\dagger}_2 b_2 b_2 + B'_{10} b_1 b_1 b_1 b^{\dagger} _3 \nonumber \\
	& + B'_{11} b_1 b_1 b_2 b^{\dagger}_4 + B'_{12} b^{\dagger}_1 b^{\dagger}_1 b^{\dagger}_1 b_3 + B'_{13} b^{\dagger}_1 b^{\dagger}_1 b^{\dagger}_2 b_4 .
	\label{eq:hamiltonian_with_scaled_op_K4}
\end{align}
For clarity, we present the indices of the monomials of $H_{K=4}$ in Table \ref{tab:coefficients_K4}, where we also list the coefficient and the corresponding circuit modules (Table \ref{tab:all_items}) for each monomial. With this, we design the quantum walk states and present the corresponding input scheme of the second-quantized many-boson Hamiltonian according to Eq. \eqref{eq:Hamiltonian_input_scheme}, where we take $\Xi = B'_4 = 29.4356$ MeV$^2$.

\begin{table}[ht]
	\caption{Information of the many-boson Hamiltonian $H_{K=4}$ for the circuit design of the Hamiltonian input scheme. The first, second, and third columns list the index $j$, the circuit modules of the combinations of the squeezed boson operators, and the coefficient of each monomial, respectively.  
	}
	\begin{tabular}{cccc|cccc} 
		\hline \hline
		index $j$ \ \ \  & circuit modules  \ \ \ &  $B'_j$ [MeV$^2$] & & & index $j$ \ \ \  & circuit modules    &  $B'_j$ [MeV$^2$] \\ \hline
		0  &  $(+-)_1$         &  $4$              & & &    7   &   $(-)_1(++)_2(-)_3$    &     $ 4.24866 $  \\
		1  &  $(+-)_2$         &  $1$              & & &    8   &   $(+)_1(--)_2(+)_3$  &    $ 4.24866 $ \\
		2  &  $(+-)_3$         &  $1/3$              & & &    9   &   $(++--)_2$        &     $1.83972 $         \\
		3  &  $(+-)_4$         &  $1/4$              & & &    10  &   $(---)_1(+)_3$      &   $ 5.66488 $     \\
		4  &  $(++--)_1$       &  $ 29.4356 $        & & &    11  &   $(--)_1(-)_2(+)_4$  &  $ 7.35889 $     \\
		5  &  $(+-)_1(+-)_2$   &  $ 29.4356 $              & & &    12  &   $(+++)_1(-)_3$       & $ 5.66488 $     \\
		6  &  $(+-)_1(+-)_3$   &  $ 9.81186 $              & & &    13  &   $(++)_1(+)_2(-)_4$        &   $ 7.35889 $  \\
		\hline \hline
	\end{tabular}
	\label{tab:coefficients_K4}
\end{table}

We note that there are five many-boson bases for our model problem with $K=4$. These bases can be sorted into two sectors according to the evenness and oddness of the particle number as $\mathcal{B}_{\rm even} = \{ \ket{3^1, 1^1}, \ket{2^2}, \ket{1^4} \}$ and  $ \mathcal{B}_{\rm odd} = \{  \ket{4^1}, \ket{2^1, 1^2} \} $. The many-boson Hamiltonian matrix in the basis $\mathcal{B}_{K=4}= \mathcal{B}_{\rm even} \oplus \mathcal{B}_{\rm odd}  $ is
\begin{align}
	H_{K=4} = \begin{pmatrix}
		 3.78630 & 1.50213 & 3.46902 &  &   \\
		 1.50213 & 1.91986 & 0 &  &  \\
		 3.46902 & 0 & 26.0767 &  &  \\
		 &  &  & 0.25 & 1.83972 \\
		 &  &  & 1.83972 & 13.5383 
	\end{pmatrix} ,
\end{align}
in units of MeV$^2$.
We find that the Hamiltonian matrix is block diagonal. This is due to the fact that the LF Hamiltonian [Eq. \eqref{eq:original_Phi4_Hamiltonian}] preserves the symmetry of oddness and evenness.
By straightforward matrix diagonalization, we obtain the eigenvalues of $H_{K=4}$ to be $E_0 = 1.61752 \times 10 ^{-7}$ MeV$^2$, $E_1 = 0.958969 $ MeV$^2$, $E_2 = 4.21772 $ MeV$^2$, $E_3 = 13.7883$ MeV$^2$, $E_4 = 26.6062 $ MeV$^2$, where $E_ 0 $ and $E_3$ are the eigenvalues for the odd-particle-number sector (lower-right submatrix), while the others are the eigenvalues for the even-particle-number sector (upper-left submatrix).

\subsection{Hybrid spectral calculations and discussions}
We adopt the SA-QKSD algorithm (Sec. \ref{sec:App_SA_QKSD}) in the eigenvalue calculations of $H_{K=4}$.
We note that our Hamiltonian input scheme respects the symmetry (the evenness and oddness) of the many-boson system.
Therefore, we can generate the Krylov bases with specific symmetry by preparing the many-boson pivot state $\ket{\psi _0}$ to be of the desired symmetry (this preparation is straightforward in the framework of second quantization) on the quantum computer.
Based on the input scheme of the Hamiltonian [Eq. \eqref{eq:Hamiltonian_input_scheme}] and that of the Chebyshev polynomial of the first kind [Eqs. \eqref{eq:Chebyshev_odd} and \eqref{eq:Chebyshev_even}], we evaluate the expectation values [Eq. \eqref{eq:expectationVals}] utilizing the noiseless Statevector simulator of the IBM Qiskit package \cite{Qiskit}. 
These expectation values are input to the classical computer to compute the Hamiltonian and the overlap matrix elements according to Eqs. \eqref{eq:H_ij} and \eqref{eq:S_ij}, which are then applied to construct the generalized eigenvalue equation to solve for the lowest eigenvalue in the SA Hilbert subspace. 

Within the Hilbert subspace of odd particle numbers, we obtain the ground-state eigenenergy to be $1.61752 \times 10^{-7} $ MeV$^2$. On the other hand, the ground state eigenvalue is $0.958969 $ MeV$^2$ when we work in the subspace of even particle numbers.
These eigenenergies agree with those from the classical calculations. The small ground-state eigenvalue in the odd-particle-number sector implies a vanishing mass gap [Eq. \eqref{eq:m2_op}]. This indicates that the coupling constant $\lambda /m^2 = 92.4746 $ is close to the critical coupling for the $(\phi ^4)_2$ theory at $K=4$. 

For the $(\phi ^4)_2$ theory in the DLCQ framework, we expect that our input scheme can face numerical challenges in the setup when we do not truncate the number of particles and when the strong coupling constant $\lambda /m^2 \gg 1$ is retained in the many-body interaction term.  However, one does not expect these challenging cases in many field theory problems of higher spatial dimensions, such as the $(d+1)$-dimensional $\phi ^4$ (with $d\geq 2$) theory, and other realistic field theory problems such as quantum electrodynamics, quantum chromodynamics, or chiral effective field theories. In such field theory problems, the coupling strengths are small, and it is reasonable to restrict the number of constituent particles in a physical system utilizing the schemes of the Fock-sector expansion and truncation within the LF Hamiltonian formalism. With such considerations, we expect that our input scheme can be adopted to treat these theories \cite{RevModPhys.21.392,BRODSKY1998299,Vary:2009gt} in a straightforward manner.

\section{Summary and outlook}
\label{sec:summary_and_outlook}

In this work, we propose a novel input scheme for second-quantized many-boson Hamiltonians in quantum computing. 
For clarity, we present the input scheme utilizing the simple two-dimensional $\phi ^4$ theory within the Hamiltonian formalism obtained by the discretized light-cone discretization approach.

In our input scheme, we employ a set of registers to encode the many-boson states, where each register encodes the occupation of a specific boson mode in binaries. 
We also squeeze the boson operators of each mode with the maximal occupation number of the mode. 
For all the monomials in the light-front (LF) many-boson Hamiltonian, we find a limited set of unique combinations of the squeezed-boson operators, and we rewrite the Hamiltonian in terms of these unique combinations. 
In this work, we present explicit design for the circuit modules (or circuit representation) for these combinations.
Based on these modules, we prepare the quantum walk states according to the many-boson Hamiltonian.
With these walk states, we block encode the many-boson Hamiltonian. 

For demonstration purposes, we present the spectral calculations of the LF Hamiltonian of the two-dimensional $\phi ^4$ theory in a restricted model space based on our Hamiltonian input scheme and implementing the hybrid quantum-classical algorithm of the symmetry-adapted quantum Krylov subspace diagonalization \cite{Du:2024zvr,Kirby_2023}. 
We perform our quantum computations utilizing the noiseless Statevector simulator of the IBM Qiskit package \cite{Qiskit}, of which the results are post-processed on the classical computer for the eigenenergies of the low-lying states.
Our results of the hybrid quantum-classical calculations agree with those exact results from direct matrix diagonalization on a classical system.

Our input scheme can be applied to treat other field theory problems, e.g., quantum electrodynamics, quantum chromodynamics, or chiral field theories [e.g., Ref. \cite{Du:2019ips}]. 
Combined with the input scheme for the second-quantized many-fermion Hamiltonians \cite{Du:2023bpw,Du:2024zvr}, our input scheme for the second-quantized many-boson Hamiltonians opens up a novel and systematic path for solving the structure and dynamics of quantum field theory on future fault-tolerant quantum computers.

\section*{Acknowledgements}
We acknowledge fruitful discussions with Peter Love, Chao Yang, Pieter Maris, Michael Kreshchuk, and Shreeram Jawadekar.
This work was supported by US DOE Grant DE-SC0023707 under the Office of Nuclear Physics Quantum Horizons program for the ``{\bf Nu}Nuclei and {\bf Ha}drons with {\bf Q}uantum computers ({\bf NuHaQ})" project.

\bibliography{apssamp}
\end{document}